\documentclass[conference]{IEEEtran}
\IEEEoverridecommandlockouts
\usepackage{cite}
\usepackage{amsmath,amssymb,amsfonts}
\usepackage{algorithmic}
\usepackage{graphicx}
\usepackage{textcomp}
\usepackage{xcolor}
\usepackage{booktabs,multirow}

\usepackage{comment}

\def\BibTeX{{\rm B\kern-.05em{\sc i\kern-.025em b}\kern-.08em
    T\kern-.1667em\lower.7ex\hbox{E}\kern-.125emX}}
\begin{document}

% REISGEN: Requirements Elicitation Interview Script Generator
% LLM-Powered Interview Script Generation for Requirements Elicitation Training
% LLM-based Automated Generation of Elicitation Interview Scripts for Requirements Engineering Education
\title{GPT-Powered Elicitation Interview Script Generator for Requirements Engineering Training

%\thanks{The second author has been partially supported by the Scientific and Technological Research Council of Türkiye through BIDEB 2232 grant no. 118C255.}
}

%\begin{comment}
\author{\IEEEauthorblockN{Binnur Görer}
\IEEEauthorblockA{
%\textit{Department of Computer Engineering} \\
\textit{Microsoft}\\
Istanbul, Türkiye \\
binnurgorer@microsoft.com}
\and
\IEEEauthorblockN{Fatma Başak Aydemir}
\IEEEauthorblockA{
%\textit{Department of Computer Engineering} \\
\textit{Utrecht University}\\
Utrecht, Netherlands \\
f.b.aydemir@uu.nl}
}
%\end{comment}

\begin{comment}
\author{\IEEEauthorblockN{\hspace{1cm}}
\IEEEauthorblockA{\textit{\hspace{1cm}} \\
\textit{\hspace{1cm}}\\
\hspace{1cm} \\
\hspace{1cm}}
\and
\IEEEauthorblockN{\hspace{1cm}}
\IEEEauthorblockA{\textit{\hspace{1cm}} \\
\textit{\hspace{1cm}}\\
\hspace{1cm} \\
\hspace{1cm}}
}
\end{comment}

\maketitle

\begin{abstract}
Elicitation interviews are the most common requirements elicitation technique, and proficiency in conducting these interviews is crucial for requirements elicitation. Traditional training methods, typically limited to textbook learning, may not sufficiently address the practical complexities of interviewing techniques. Practical training with various interview scenarios is important for understanding how to apply theoretical knowledge in real-world contexts. However, there is a shortage of educational interview material, as creating interview scripts requires both technical expertise and creativity. To address this issue, we develop a specialized GPT agent for auto-generating interview scripts. The GPT agent is equipped with a dedicated knowledge base tailored to the guidelines and best practices of requirements elicitation interview procedures. We employ a prompt chaining approach to mitigate the output length constraint of GPT to be able to generate thorough and detailed interview scripts. This involves dividing the interview into sections and crafting distinct prompts for each, allowing for the generation of complete content for each section. The generated scripts are assessed through standard natural language generation evaluation metrics and an expert judgment study, confirming their applicability in requirements engineering training.
\end{abstract}


\section{Introduction}
\label{sec:introduction}

% REI as the most favored method, and its strengths
Requirements elicitation stands as a pivotal stage in the software development lifecycle, focusing on grasping and documenting the needs and expectations of stakeholders. This phase is essential for ensuring that the final product aligns with the intended purpose and user requirements. Among the various techniques utilized for requirements elicitation, interviews emerge as the foremost and most prevalent method~\cite{zowghi2005requirements}. The strength of interviews lies in their interactive nature, permitting analysts to directly engage with stakeholders. This direct interaction facilitates a comprehensive gathering of crucial information, aids in the clarification of requirements, and fosters a mutual understanding of the envisioned system. Engaging in interviews ensures a thorough and nuanced comprehension of stakeholder needs, which is vital for the successful development of software that meets user expectations. However, conducting a successful interview requires a set of complex skills that can be particularly challenging for novice analysts to acquire and implement effectively.

% REI training
Traditional training methods, typically limited to textbook learning, may not sufficiently convey the complexities of interview techniques and how to apply what's learned in the books to real-life situations. On the other hand, organizing training interviews with actual stakeholders in a real-world domain presents significant logistical challenges, particularly regarding stakeholder availability and the practicality of managing large numbers of students. To address this limitation, role-playing has emerged as the most prevalent instructional technique~\cite{ferrari2020sapeer}. In this method, students alternate between playing the roles of the stakeholder and the analyst. This approach allows them to experience firsthand the challenges of conducting interviews from both perspectives. However, role-playing is also resource-intensive; organizing these sessions, coupled with the subsequent analysis of the interviews to provide feedback, is time-consuming for educators. This inherently limits the number of interview experiences students can engage in, impacting the depth and diversity of their practical training.

% Our proposal is to address the issues of REI training with gen AI 
An alternative method to enhance students' practical comprehension of interview processes could involve the provision of interview scripts.
This approach would allow students to study and analyze structured examples of how interviews are conducted, offering them a tangible reference point. By reviewing these scripts, students can gain insights into the flow of questions, the nuances of interaction, and the complexity of eliciting information. This method not only aids in understanding the theoretical aspects of interviews but also provides a more hands-on learning experience, bridging the gap between theory and practice in a more interactive and engaging manner. These scripts can be versatile in their application, functioning as valuable course materials, illustrative tools for classroom discussions, or as resources for independent student study.

However, to the best of our knowledge, no dataset is available that includes elicitation interview scripts from real or simulated interviews. Obtaining scripts from real interviews poses significant challenges, as they may contain company confidential information like trade secrets, or the participants involved might not consent to collect their interviews for educational purposes. Simulated interviews, often conducted within research studies for specific research objectives, are either not recorded or, if they are, the recordings are not made publicly available~\cite{bano2019teaching, dalpiaz2021deriving, ferrari2022requirements}. In this context, the potential of artificial intelligence (AI)-generated scripts emerges as a viable alternative. The field of generative AI, particularly the advancements in large language models (LLMs), opens up new frontiers and presents unique opportunities in educational content creation~\cite{sarsa2022automatic, bulathwela2023scalable}. These advanced AI models offer the potential to generate realistic and varied interview scripts, presenting an opportunity to contribute to the supplementary educational materials in elicitation interview training while reducing the human effort involved.

In our previous work~\cite{gorer2023generating}, we explore prompt engineering to create interview scripts with intentional mistakes typical in requirements elicitation interviews. Our methodology is designed to showcase the potential errors an analyst might make in their questioning, thereby providing students with practical examples of incorrect interviewing techniques.

In this paper, we introduce a GPT-powered interview script generator for employing LLMs in generating mistake-free and lengthy interview scripts, addressing the limitations of the state-of-the-art, especially regarding the input and output context length of LLMs. We validate the applicability of these scripts through both quantitative and qualitative evaluations. Our contributions are two-fold:
\begin{enumerate}
    \item We introduce a GPT-powered interview script generator with a dedicated knowledge base tailored to the well-established guidelines and best practices of requirements elicitation interviews.
    We employ a prompt chaining approach to mitigate the output length constraint of GPT to generate thorough and detailed interview scripts. This involves dividing the interview into separate sections and crafting distinct prompts for each, allowing for the generation of complete content for each interview section.
    \item We assess the generated scripts through standard natural language generation (NLG) evaluation metrics specifically used for dialogue tasks and an expert judgment study, confirming their applicability in requirements engineering training. 
\end{enumerate}

The remainder of this paper is organized as follows: 
%Section~\ref{sec:background} gives the background on elicitation interview scripts and large language models.
Section~\ref{sec:related_work} provides an overview of related work on the use of large language models in software engineering and software engineering education. 
Section~\ref{sec:approach} describes our proposed approach for generating interview scripts, including configuration of GPT and prompt engineering techniques employed. 
Section~\ref{sec:evaluation} presents the results of automatic evaluations and human judgment study, along with discussions of the findings. 
Section~\ref{sec:research_plan} outlines our future research agenda and Section~\ref{sec:conclusion} concludes the paper.

%\section{Background}
%\label{sec:background}
%\subsection{Characteristics of Requirements Elicitation Interviews}
%\subsection{Natural Language Generation}

\section{Related Work}
\label{sec:related_work}
LLMs have revolutionized numerous fields, including software engineering~\cite{lo2023trustworthy}. Initially designed for natural language processing tasks, these models have demonstrated remarkable versatility in understanding programming languages and generating code. A significant area of focus has been the generation and customization of code, highlighting how LLMs can automate routine coding tasks, thereby accelerating development processes and reducing human error~\cite{hou2023large}. These models are trained on vast code repositories, enabling them to generate code snippets, suggest code improvements, and even write entire functions based on natural language prompts. Another key application is in automated test case generation and debugging~\cite{siddiq2023exploring}. LLMs can understand the context of a given codebase and generate relevant test cases, a task that is traditionally time-consuming and prone to human oversight.

LLMs are also being investigated for their use in requirements engineering and design~\cite{fan2023large}, although they have received less attention than other software engineering processes like code generation and testing~\cite{hou2023large}. Kanuka \textit{et al.} show how ChatGPT can effectively integrate specific requirements into design models and code, addressing the traceability issue between these elements~\cite{kanuka2023exploring}. Ronanki \textit{et al.} investigate ChatGPT's potential to assist in requirements elicitation processes and find that requirements generated by ChatGPT were deemed acceptable by experts, comparable in quality to those crafted by professionals~\cite{ronanki2023investigating}. Zhang \textit{et al.} offered a positive outlook on using LLMs for efficient requirements engineering after evaluating ChatGPT's performance in requirements analysis tasks~\cite{zhang2023preliminary}. Luo \textit{et al.} used BERT for automated requirement classification through prompt engineering~\cite{luo2022prcbert}, while Luitel \textit{et al.} employed BERT to enhance requirements completeness by predicting missing information~\cite{luitel2023improving}. In their detailed analysis, Arora \textit{et al.} provide SWOT analyses on the use of LLMs for each requirements engineering step, suggesting a generally optimistic yet cautious perspective of AI's role in the field~\cite{arora2023advancing}.

Integrating generative AI into software development will likely shift the required skill set for software engineers~\cite{daun2023chatgpt}. With generative AI's emerging capabilities in code development and testing, software engineers might need to focus more on conceptualization, requirements elicitation, and software design to manage and integrate AI-generated code within larger projects effectively. This evolution in the industry underscores the need for software engineering education to adapt, placing a stronger emphasis on these aspects of the software development process. Concurrently, LLMs are emerging as powerful tools in revolutionizing educational methodologies~\cite{duc2023generative}. In software engineering education, pre-trained models such as ChatGPT and Codex have been predominantly utilized as tools for generating educational content and as assistive systems. For content creation, recent research efforts have focused on tasks closely related to coding, including the development of programming exercises~\cite{macneil2023experiences} and the generation of code explanations~\cite{sarsa2022automatic}. As assistive systems, LLMs have been explored for various applications like aiding pair-programming learning processes~\cite{banic2023pair}, question-answering for software testing~\cite{jalil2023chatgpt}, and assisting with coding-related queries~\cite{savelka2023large}. The consensus from these studies is that LLMs can mostly produce accurate, novel, and useful content and answers, but the quality of these outputs can be improved by employing effective prompting techniques and including additional information in the queries. Our study builds upon existing research by investigating how pre-trained LLMs such as GPT-4 can be used to generate requirements elicitation interview scripts. This area of software education remains largely unexplored, especially in the context of applying generative AI. Our research focuses on exploring and uncovering its potential applications and benefits for requirements engineering education.

%This dual impact of generative AI – both in the professional competencies of software engineers and the pedagogical methods in education – reflects a significant transformation in the software engineering education.

%LLM-based Education Content Generation
%~\cite{kasneci2023chatgpt,dijkstra2022reading,gabajiwala2022quiz}
%\cite{jiao2023automatic}
%Automatic education question generation with difficulty level controls.
%\cite{pal2022weakly}
%Weakly Supervised Context-based Interview Question Generation, DL ML job interview question generation

% LLMs for SE research
%Beyond practical implementations, LLMs are being explored to both conduct and assist in qualitative and quantitative software engineering research studies. Liang \textit{et al.} investigates how LLMs can replicate and extend software engineering research, particularly in analyzing methodologies and automating the replication of studies~\cite{liang2023can}. On the qualitative side, Bano \textit{et al.} discusses the use of LLMs in qualitative research within software engineering, exploring their potential in synthesizing and analyzing qualitative data, a task traditionally requiring significant human effort~\cite{bano2024large}.

\section{Approach}
\label{sec:approach}  
LLMs can possess an extensive repository of general information, gained from extensive pre-training on substantial datasets~\cite{hu2023survey}. This allows them to display a broad understanding across a range of fields. Yet, this comes with notable constraints: it tends to be generic and might not delve into the finer details of specialized areas. Adapting an LLM to a particular domain requires refining and expanding the model's knowledge base with the specific data of that domain~\cite{singhal2023large,wu2023bloomberggpt}. 

Two prevalent approaches for injecting specific knowledge into a pre-trained model include fine-tuning and in-context learning (ICL)~\cite{ovadia2023fine}. Fine-tuning extends the existing model's training using specialized task-specific data, effectively calibrating the model's weights to synchronize with a particular knowledge base, thereby optimizing its performance for specific applications~\cite{dodge2020fine}. In contrast, ICL aims to improve pre-trained language models' performance on new tasks by modifying the input query rather than directly changing the model's weights~\cite{chen2021meta}. One prominent method within ICL is retrieval-augmented generation (RAG)~\cite{lewis2020retrieval}. RAG employs information retrieval techniques to pull relevant information from a knowledge source for the generated text. This approach allows customizing the model's knowledge base for the domain-specific task without modifying the model's underlying structure.

Requirements elicitation interviews, while mirroring standard interviews in having a clear goal and a question-answer format, also possess specific trends and aspects. The content and flow of an interview can vary significantly based on factors like the stakeholder's profile, the amount of information gathered before the interview, and the product or the domain itself. Still, there are common guidelines and best practices recommended in the literature to conduct a successful interview, effective across various elicitation scenarios~\cite{mohedas2022use}. For example, analysts should meticulously craft questions to create a mutual understanding of the product between them and stakeholders~\cite{donoghue2000projective} and their questions should be comprehensive enough to cover all critical facets of the product~\cite{burnay2014stakeholders}. An LLM specifically enhanced with such guidelines for requirements elicitation interviews would generate more appropriate and relevant interview scripts compared to a generic LLM~\cite{ovadia2023fine}. Fine-tuning is not a feasible option for our task as it requires a significant amount of task-specific data but there is not such a dataset for elicitation interview scripts. Hence, we opt for a RAG-like approach for knowledge injection.

Our methodology comprises two primary components: initially, we create and configure a custom GPT-4 agent, tailoring it for the specific task of generating interview scripts. This involves integrating relevant knowledge sources and instructions that align with the requirements of the task. The second part of our approach utilizes a technique known as prompt chaining. This method involves sequentially generating portions of the interview script. By doing so, we effectively circumvent the output length limitations inherent in LLMs. This two-pronged approach allows us to produce relevant and comprehensive interview scripts, overcoming the typical constraints of a standard LLM for interview script generation task. The knowledge files and links to prompt chains are publicly available\footnote{https://doi.org/10.6084/m9.figshare.25193657}.
 
\subsection{Configuring Custom GPT}
\label{sec:customgpt}
\begin{figure}[htbp]
\centering
\includegraphics[width=0.99\linewidth]{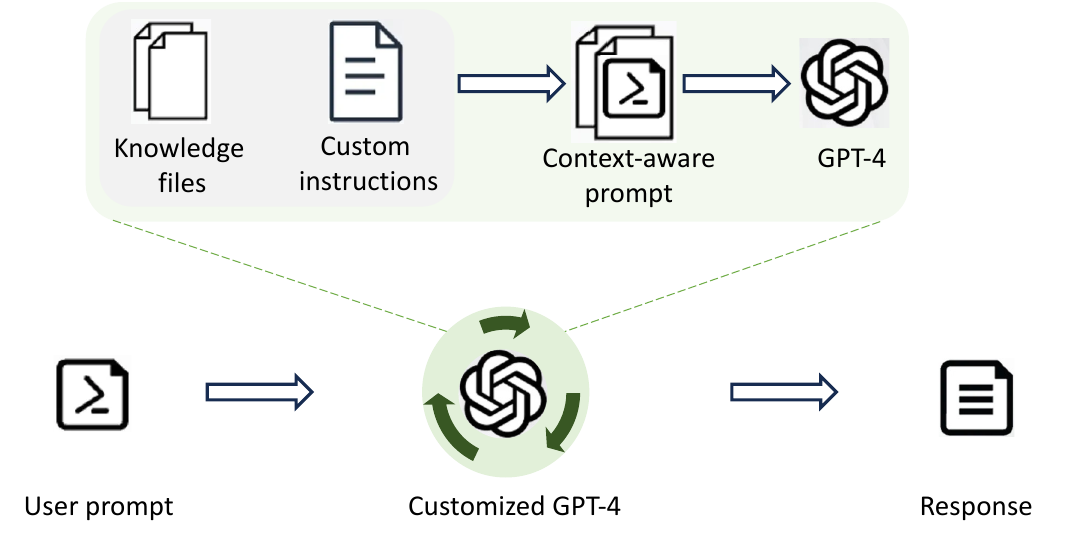}
\caption{Configuring custom GPT.}
\label{fig:flowchart_1}
\end{figure}

GPTs are custom versions of ChatGPT, offering users the capability to customize the model for specific tasks or topics by integrating tailored instructions and knowledge sources. These customized GPTs can vary in complexity to suit various roles, ranging from coding assistant to technical support agent. Custom GPTs inject the customized instructions and knowledge directly into the model's context. This means that each time a prompt is submitted, the model accesses and utilizes this tailored context--comprising specific instructions and relevant knowledge. Consequently, the model can respond to each prompt in a manner that is informed by and aligned with the customized parameters, ensuring responses are both relevant and appropriate for the task or topic at hand. Figure~\ref{fig:flowchart_1} shows the response generation flow of custom GPT.

We provide three knowledge sources to the custom GPT:

\paragraph{General guidelines for interviewer's questions} This resource equips the GPT with a methodical strategy for crafting questions. It stresses the need for questions to be clear and direct, which helps elicit comprehensive responses that encompass every aspect of the system or product. The guidelines emphasize the importance of adapting the conversation to the complexity of the product and the stakeholder's inputs, maintaining flexibility for probing deeper when necessary. This includes crafting follow-up questions to stakeholders' ambiguous answers or diving deeper into discussions for complex features. This resource also instructs on ensuring smooth transitions between different discussion points.
%domain exploration, system as-is, system to-be
\paragraph{Common pitfalls to avoid for interviewer's questions} This resource provides the GPT with knowledge of the typical errors identified in elicitation interviews, as outlined in~\cite{bano2019teaching}. By highlighting what not to do for the generation of the interviewer's questions — such as leading the stakeholder, using technical jargon, or lack of clarity — the GPT is guided to avoid these mistakes in its generated scripts, leading to more effective and error-free elicitation interviews.
\paragraph{Interview script sample} This resource is to enhance the GPT's understanding of the typical structure and dynamics of an elicitation interview script. We used the script provided in~\cite{spijkman2023summarization}, which details a simulated interaction between two analysts and a stakeholder regarding the development of a portal for the International Football Association. In the simulation, students assumed the roles of analysts, while a researcher acted as the stakeholder. To align with our focus on one-to-one conversations between an analyst and a stakeholder, we have tailored the script to feature just one analyst. This simplification was accomplished by combining the responses from the original two analysts into a single thread or by transferring the lines of the second analyst to the first. This creates a more direct and streamlined narrative for the GPT to assess and emulate in the interview generation process. 
%The interview focuses on a global portal to facilitate the organization of football leagues. 
The conversation spans 114 exchanges between the analyst and the stakeholder, starting from initial greetings and progressing through the exploration of key features, like a budgeting portal and a scheduling portal. It delves into the design aspects of these features and the system's scalability. Furthermore, the conversation encompasses inquiries about the project timeline and the identification of other stakeholders, providing a holistic view of the requirements elicitation interviews to the GPT. The average length of the analyst's responses is 37 words per turn. There are five instances where the analyst's turns are notably brief, comprising less than six words each. In contrast, the stakeholder's responses are slightly longer, averaging 49.3 words per turn. Similar to the analyst, the stakeholder also has brief moments in the conversation, with seven instances where their responses are less than six words.

\subsection{Outline-based Interview Generation}
\label{sec:segmenting_generation}
%https://platform.openai.com/docs/models/gpt-4-and-gpt-4-turbo
%Break down your goal into chunks: If you need ChatGPT to write a detailed essay, story, or code, consider dividing the task into individual subheadings or chapters. You can then ask ChatGPT to generate them one at a time. For example, you can request an introduction to the topic in one prompt, then continue for each section until you reach the conclusion.
%Ask for an outline: If you're struggling to break down the task into smaller chunks yourself, don’t forget that you can also ask ChatGPT to do it for you. In the first prompt, provide a title for the essay or story you have in mind along with any other context you need to include. Then, ask the chatbot to write each section one by one.
Using GPT to create lengthy interview scripts may raise a network error. This issue stems from GPT's output limit, a common safeguard in LLMs designed to prevent outputs from becoming repetitive or contradictory with excessive text generation~\cite{openai2023gpt4, park2023longstory}. It is advisable to avoid pushing these limits. Generating content within large limits can lead the model in unexpected directions. Producing content in shorter segments is typically more effective than in a single run.

\begin{figure*}[htbp]
\centerline{\includegraphics[width=0.75\textwidth]{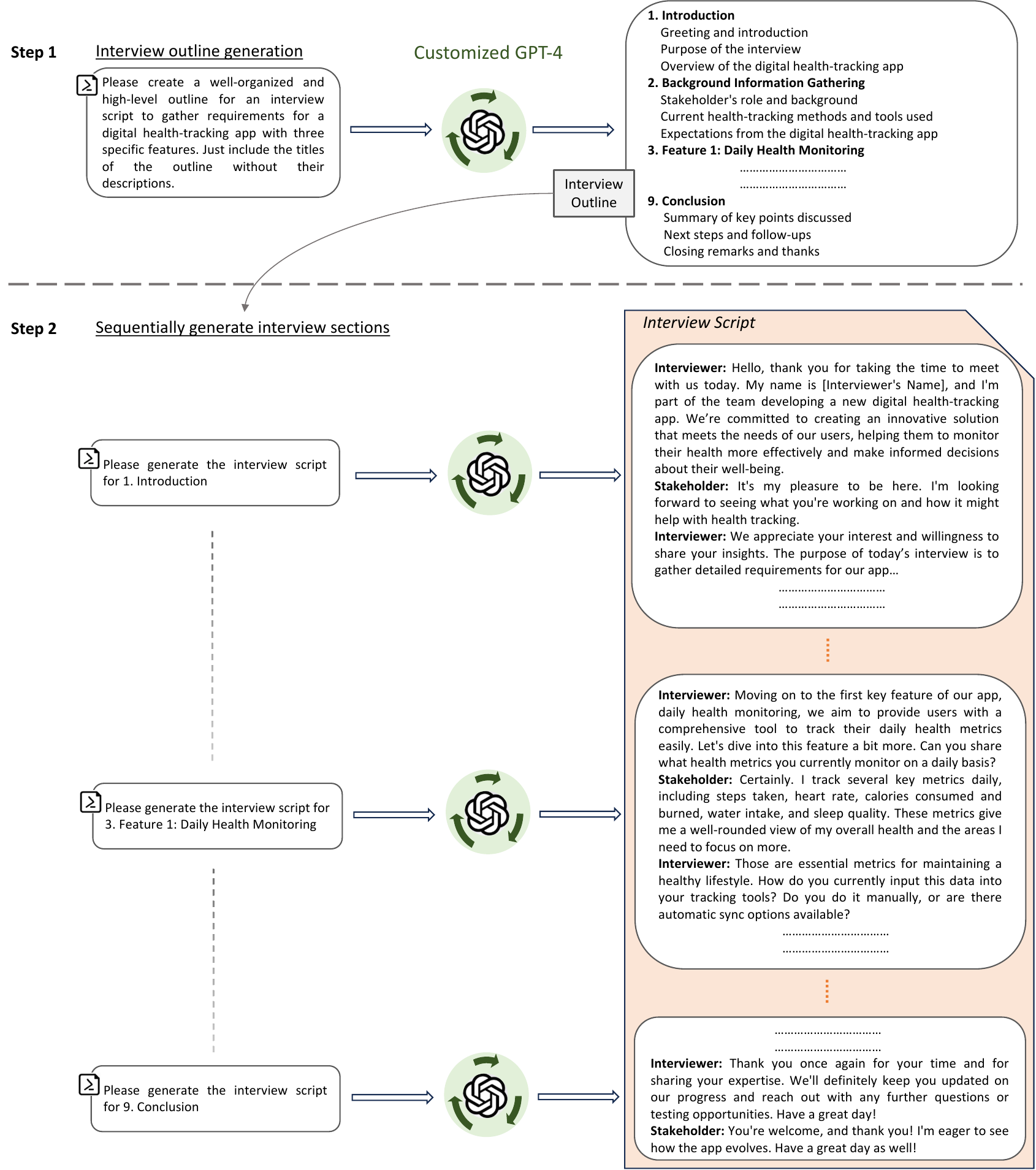}}
\caption{Outline-based interview generation with custom GPT.}
\label{fig:outlinebased_gen}
\end{figure*}

To address the challenge of creating detailed and lengthy interview scripts within these constraints, our approach employs a strategy like outline-based generation used in neural story generation~\cite{fan2018hierarchical, rashkin2020plotmachines, yang2022re3, yang2022doc}. This involves initiating the process with a primary prompt that outlines the interview's various sections. We then systematically generate each section with tailored prompts, ensuring focused and coherent development. This multi-prompt approach ensures dedicated attention to each part of the interview, thereby promoting the creation of detailed and coherent sections. Once all prompts have been executed and respective sections generated, we concatenate them into a singular, comprehensive interview script. This method not only effectively navigates the token limit restriction but also enhances the overall coherence of the final script, as each section is generated in alignment with the interview outline determined by the initial prompt. Figure~\ref{fig:outlinebased_gen} illustrates the outline-based interview generation.

% zhang2023survey
%A survey of controllable text generation using pre-trained language models
%Controlled elements-> Generative Model based on pre-trained language models (PMLs) -> Generated text
%Table 1, dialogue generation, controllable aspect(persona, politeness, sentiment, template, ground-truth reference)

\section{Evaluation}
\label{sec:evaluation}
To assess the effectiveness of our methodology, we developed four distinct interview scripts, each tailored to a specific domain: a meeting scheduler system (S1), a social housing application (S2), a digital health tracking application (S3), and a food delivery application (S4). The generated interview scripts are publicly available\footnote{https://doi.org/10.6084/m9.figshare.25193657}. We compiled dialogue characteristics for each generated interview script, which are given in Table~\ref{tab:stats}. For both interviewer and stakeholder turns, the statistics for the length of turns in words are presented, including minimum to maximum values, first quartile ($Q_1$), median ($Mdn$), and third quartile ($Q_3$), as well as the number of dialogue acts categorized into non-question (NQ) and question (Q) types. The top 10 terms with the highest TF-IDF (Term Frequency-Inverse Document Frequency) scores for each scenario are also provided, indicating key thematic elements discussed. Interviewer turns show variability in length across scenarios, indicating a difference in the detail or complexity of questions asked by interviewers or the responses provided by stakeholders. Stakeholder turns, on average, tend to be longer in S2 and S3, indicating more detailed responses or more complex interviewer questions that require longer answers in these scenarios. The number of questions (Q) and non-questions (NQ) varies significantly, with S3 showing a notably high number of questions from the interviewer, suggesting a more inquisitive approach in this scenario compared to others.
Stakeholder turns do not include questions in any scenario, which is typically expected as the stakeholders primarily respond to the interviewer's queries. The high-frequency terms per script align with the thematic focus of each scenario.

\begin{table*}[htbp]
\caption{Dialogue characteristics of each generated interview script.}
\label{tab:stats}
\centering
\begin{tabular}{|l|cccccc|cccccc|l|}
\hline
                       & \multicolumn{6}{c|}{\textit{Interviewer Turns}}                                                                              & \multicolumn{6}{c|}{\textit{Stakeholder Turns}}                                                                              & \multicolumn{1}{c|}{\multirow{3}{*}{\begin{tabular}[c]{@{}c@{}}Top 10 Terms with highest \\ TF-IDF scores\end{tabular}}}                                   \\ \cline{2-13}
\multicolumn{1}{|c|}{} & \multicolumn{4}{c|}{Length (in words)}       & \multicolumn{2}{c|}{\begin{tabular}[c]{@{}c@{}}Dialogue \\ Acts\end{tabular}} & \multicolumn{4}{c|}{Length (in words)}       & \multicolumn{2}{c|}{\begin{tabular}[c]{@{}c@{}}Dialogue \\ Acts\end{tabular}} & \multicolumn{1}{c|}{}                                                                                                                                      \\ \cline{2-13}
\multicolumn{1}{|r|}{} & $\min\text{-}\max$ & $Q_1$ & $Mdn$ & \multicolumn{1}{c|}{$Q_3$} & NQ                                    & Q                                     & $\min\text{-}\max$ & $Q_1$ & $Mdn$ & \multicolumn{1}{c|}{$Q_3$} & NQ                                     & Q                                    & \multicolumn{1}{c|}{}                                                                                                                                      \\ \hline
\textbf{S1}            & 3-100   & 28 & 36  & \multicolumn{1}{c|}{49} & 10                                    & 25                                    & 10-44   & 23 & 28  & \multicolumn{1}{c|}{38} & 34                                     & 0                                    & \begin{tabular}[c]{@{}l@{}}scheduling, scheduler, preferences, \\ priorities, calendar, integration, \\ automated, user, rescheduling\end{tabular}         \\ \hline
\textbf{S2}            & 9-144   & 22 & 33  & \multicolumn{1}{c|}{46} & 8                                     & 25                                    & 2-84    & 38 & 50  & \multicolumn{1}{c|}{55} & 33                                     & 0                                    & \begin{tabular}[c]{@{}l@{}}applicants, notifications, tenants, \\ eligibility, rent, maintenance, dashboard,\\ requests, streamline, envision\end{tabular} \\ \hline
\textbf{S3}            & 14-84   & 25 & 33  & \multicolumn{1}{c|}{41} & 11                                    & 44                                    & 17-72   & 39 & 46  & \multicolumn{1}{c|}{52} & 55                                     & 0                                    & \begin{tabular}[c]{@{}l@{}}tracking, metrics, personalized, feedback,\\ recommendations, privacy, user, data, \\ health, insights\end{tabular}             \\ \hline
\textbf{S4}            & 2-135   & 35 & 40  & \multicolumn{1}{c|}{48} & 1                                     & 26                                    & 4-70    & 21 & 35  & \multicolumn{1}{c|}{47} & 27                                     & 0                                    & \begin{tabular}[c]{@{}l@{}}menu, delivery, envision, user, browsing,\\ ordering, considerations, functionalities,\\ dietary, tracking\end{tabular}         \\ \hline
\end{tabular}
\end{table*}

Evaluating the quality of AI-generated text is complex, given that ``quality" itself can mean different things in different contexts, rendering a one-size-fits-all metric unfeasible. Human evaluation, with its nuanced understanding of quality, is considered the golden standard. However, its use is constrained by the considerable time and cost involved. Consequently, automated evaluation methods have gained popularity as a more practical alternative. Our approach integrates an automated evaluation method, which inspects common linguistic quality attributes, with an expert judgment study that places a greater emphasis on metrics specific to the quality of requirements engineering interviews.

\subsection{Automated Evaluation}
\label{sec:auto_evaluation}
%Rule-based metrics, Machine-learned metrics, LLM-based metrics ~\cite{gong2023coascore, liu2023gpteval}
%LLM-based metrics ~\cite{ke2023decompeval, li2024leveraging}
Automated evaluation metrics for NLG are generally categorized into two main types: reference-based and reference-free~\cite{celikyilmaz2020evaluation}. Reference-based metrics evaluate the quality of generated text by comparing it to one or more ground-truth references, focusing primarily on the similarity of n-gram sequences between the generated text and reference materials. In contrast, reference-free evaluations assess the generated text on its own merits, evaluating semantic quality and grammatical correctness without comparing it to pre-existing texts. These evaluations often rely on neural-based models to gauge the text's coherence and fluency. Given the lack of ground-truth references in our study, we chose a reference-free approach for evaluation. Specifically, we used the GRUEN metric developed by Zhu \textit{et al.}~\cite{zhu2020gruen}, which has proven effective in evaluating multiple NLG tasks, including dialogue generation. Dialogue generation closely aligns with our objective of developing interview scripts, making GRUEN a fitting choice for our evaluation needs, as opposed to other metrics tailored to specific NLG tasks such as text summarization or data augmentation~\cite{celikyilmaz2020evaluation}.

GRUEN evaluates text based on four key criteria: grammaticality, non-redundancy, focus, and coherence. A high grammaticality score indicates that the system's output is readable, fluent, and free of grammatical errors. Non-redundancy emphasizes the avoidance of unnecessary repetition, such as using specific names repetitively when pronouns would suffice. Focus is about ensuring that there is a semantic connection between neighboring sentences, making the text cohesive. Lastly, coherence refers to the text's organization, where sentences are arranged in a natural, logical order that is easy for readers to follow. GRUEN scores range from 0 to 1, with higher values indicating better language quality. We calculate the GRUEN scores for each interview turn. Table~\ref{tab:gruen_results} presents these scores, detailing the averages and standard deviations for both interviewer and stakeholder turns, as well as a combined average for all turns, across each interview script. The results indicate a generally high linguistic quality in the dialogues across all scenarios, with minor variations between interviewer and stakeholder turns. 

\begin{table}[htbp]
\centering
\caption{GRUEN scores for each generated interview script.}
\label{tab:gruen_results}
\begin{tabular}{lccc}
\toprule
   & \begin{tabular}[c]{@{}c@{}}Interviewer Turns\\ ($mean \pm std$)\end{tabular} & \begin{tabular}[c]{@{}c@{}}Stakeholder Turns\\ ($mean \pm std$)\end{tabular} & \begin{tabular}[c]{@{}c@{}}All Turns\\ ($mean \pm std$)\end{tabular} \\ 
   \midrule
\textbf{S1} & 0.81$\pm$0.05                                                              & 0.77$\pm$0.05                                                              & 0.79$\pm$0.06                                                          \\ 
\textbf{S2} & 0.79$\pm$0.05                                                              & 0.82$\pm$0.03                                                              & 0.81$\pm$0.04                                                          \\ 
\textbf{S3} & 0.82$\pm$0.04                                                              & 0.79$\pm$0.05                                                              & 0.81$\pm$0.05                                                          \\ 
\textbf{S4} & 0.80$\pm$0.05                                                              & 0.79$\pm$0.05                                                              & 0.79$\pm$0.05                                                          \\ \bottomrule
\end{tabular}
\end{table}

\subsection{Expert Judgment Study}
\label{sec:expert_study}
We conducted a judgment study with an expert who is an active researcher in the community, has been teaching requirements engineering courses for over seven years, and has been collaborating with the industry for over five years. 

The expert evaluated the generated scripts from different views. The first view is the natural language view, where we identified three dimensions:
\begin{itemize}
\item Naturalness (natural flow): This dimension assesses whether the responses feel natural within the context of the dialogue history. The expert found the generated scripts mostly natural, with few exceptions. For example, the interviewer mentions ``the next section" which is unnatural for spoken language. Also, the stakeholder is very cooperative in all cases, which may not always be the case in an interview. 
\item Coherence: This dimension assesses how well the parts of a text fit together logically and orderly, ensuring that the content flows smoothly and is understandable to the reader. The expert confirms that the generated scripts are coherent, the dialogue turns flow meaningfully one after the other, and the content is logically linked. 
\item Completeness: This measure evaluates whether the generated text covers all necessary aspects of the topic it aims to address, ensuring that all relevant information is presented and no critical details are omitted. This is the dimension where the expert identifies the most room for improvement. The expert noted that the overall elicitation process should cover the domain, stakeholders, current business processes, the as-is system, and the needs and desires of the to-be system. However, even in real-life elicitation sessions, covering them all in a single session may not be possible. A session may focus on gathering specific types of information, such as needs and desires for the future system. According to the expert, the generated scripts briefly touch on all these topics but lack depth. This shortfall is particularly evident for non-functional requirements. For instance, while privacy issues are discussed in the context of health tracking application (S3), fairness is overlooked in the social housing project scenario (S2).
\end{itemize}
Table~\ref{tab:expert_nlg} presents the expert's evaluation scores for each dimension across various scenarios. The uniformity of these scores, irrespective of the domain, suggests that the interview script generator consistently produces natural and coherent scripts. However, this consistency also highlights that the generator needs more effective prompting to achieve completeness in its outputs, particularly in the depth of discussion on the topics covered.

\begin{table}[htbp]
\centering
\caption{Expert evaluation results for natural language quality attributes of each generated interview script.}
\label{tab:expert_nlg}
\begin{tabular}{p{1.4cm}ccllll}
\toprule
\multirow{2}{=}{\textbf{Dimension}}& \multicolumn{2}{c}{\textbf{Scale}}& \multicolumn{4}{c}{\textbf{Score}}\\
\cmidrule(lr){4-7}
           &  \textbf{1 ...} & \textbf{... 5}     & \textbf{S1}& \textbf{S2}& \textbf{S3} & \textbf{S4} \\ \midrule
Naturalness & \begin{tabular}[c]{@{}c@{}}not natural\\at all\end{tabular}  & \begin{tabular}[c]{@{}c@{}}quite\\natural\end{tabular}                       &   4       &4&    4 & 4\\ 
Coherence & \begin{tabular}[c]{@{}c@{}}not coherent\\at all\end{tabular} & \begin{tabular}[c]{@{}c@{}}quite\\coherent\end{tabular}      &  4         &4&   4 & 4\\ 
Completeness & \begin{tabular}[c]{@{}c@{}}not complete\\at all\end{tabular} & \begin{tabular}[c]{@{}c@{}}quite\\complete\end{tabular}  &   3          &3&  3 &3\\
\bottomrule
\end{tabular}
\end{table}

For the second view, the expert assessed the scripts using the requirements elicitation interview rubric introduced in~\cite{lending2022rubric}. This rubric covers the essential elements of an effective requirements elicitation interviewing process. Table~\ref{tbl:rubric} presents the rubric elements and the expert evaluation scores ranging from 1 (lowest score) to 5 (highest score) for each element across all the scripts. The expert's feedback highlights the following key observations across the interview scripts. In all scripts, the interviewer builds rapport at the beginning of the interview; however, the stakeholder's relationship with the project is not always clearly identified. For script S3, we understand the stakeholder is an end-user, and in script S2, the stakeholder is a customer, but the role of the stakeholder within the organization is not specified. The features discussed for the design of ``to-be" system are reasonable and related to the project, but they seem to be set before the interview, and no new discoveries unfold during the interview. At the end of each interview, there is a summary that is not in-depth but still provides an overview of the discussion. The good practice of getting the approval of the stakeholder is present in all scripts. Regarding active listening, the interviewer follows the response of the stakeholder but does not always achieve depth based on the response. For example, in all the domains, the stakeholder highlights the importance of privacy, yet the interviewer does not further probe into how these privacy concerns could be addressed in their subsequent responses.

\begin{table}[htbp]
\centering
\caption{Expert evaluation results on the requirements elicitation interview rubric from ~\cite{lending2022rubric}.}
\label{tbl:rubric}
\begin{tabular}{p{4.5cm}llll}
\toprule
\multirow{2}{=}{\textbf{Rubric}}& \multicolumn{4}{c}{\textbf{Score}}\\
\cmidrule(lr){2-5}
                  & \textbf{S1}& \textbf{S2}& \textbf{S3} & \textbf{S4} \\ \midrule
Greeting                           &   2       &3&    3 & 3\\ 
Opening      &  3         &3&   3 & 4\\ 
Analyze Current State ``As Is"              &   3          &3&  3 &3\\ 
Design ``To Be" System &    3            &3&3&3\\ 
Closing                &    4           &4& 4&3\\ 
Active Listening                    &    2            &3&3&3\\ 
\bottomrule
\end{tabular}
\end{table}

For the third view, the expert analyzed the interviews concerning the most common interview mistakes~\cite{bano2019teaching}. Among the list of the most common mistakes, \textit{influencing stakeholder}, \textit{unnatural dialogue style}, and \textit{ignoring other stakeholders} are mostly present in the scripts. In the interviews, there is a notable instance of \textit{influencing stakeholder}, where the features to be discussed seem predetermined at the start of the interview instead of being organically explored and identified throughout the conversation. \textit{Unnatural dialogue style} mostly because of the having written language in the script generation, making it unnatural for spoken language. \textit{Ignoring other stakeholders} is the most prominent mistake, as the interviewer does not ask for additional stakeholders in any of the scripts.

\section{Research Plan}
\label{sec:research_plan}
In this section we list our plan to continue our research. More extensive evaluation of our existing work, involving multiple LLMs for the interviewer and the interviewees, additional use cases for our research and their evaluation are parts of our research plan. 

 We aim to conduct studies on the perceived usefulness and effectiveness of scripts as an educational tool with students who take our graduate requirements engineering courses to receive their perceptions on the scripts as an educational material. We will also increase the number of experts for our judgment study by reaching out to the members of the requirements engineering community. %More evaluations as there are concerns on if we can trust AI-generated educational content ~\cite{nguyen2023generative} mentions positive claims ~\cite{sallou2023breaking} ~\cite{bull2023generative}

The primary limitation of the existing methodology lies in its ineffectiveness in uncovering new requirements throughout the dialogue. To address this challenge, we intend to refine our knowledge base with updated guidelines for interviewers, particularly emphasizing the importance of exploring and identifying new requirements during conversations. Additionally, we propose to develop a series of interviews, each dedicated to a distinct phase of the elicitation process: domain discovery, analysis of the current (``as-is") system, and envisioning the future (``to-be") system. This structured approach aims to produce more detailed and focused scripts for each stage.

In our current research a single LLM generates the whole script. We plan to configure separate LLMs as the interviewer and the interviewee to interact and generate the interviewer and the stakeholder scripts, allowing us further customize the interviewer and interviewee with persona creation for stakeholders with various personality traits, job status, or technology familiarity~\cite{lee2022personachatgen}.

A potential use case of our research is to support to elicitation process by providing additional stakeholders where the availability of certain type of the stakeholders is low. We plan to evaluate the outputs of the interviewer and interviewee LLMs in experimental simulations and in field studies~\cite{10.1145/3241743}, assessing their effectiveness and comparing their performance with humans.

\section{Conclusions}
\label{sec:conclusion}
Our study highlights the lack of practical training materials for conducting interviews in requirements elicitation, along with the difficulty of creating these materials, especially when it comes to sample interview scripts. To address this, we developed an LLM-based solution: a GPT agent designed to generate interview scripts automatically. This agent, built with a knowledge base grounded in the core principles of requirements elicitation interviews, leverages a prompt chaining technique to overcome LLMs' limitations on output length, enabling the production of comprehensive scripts. We demonstrated the effectiveness of the generated scripts, using both computer-based methods to assess the language quality and expert judgment evaluations to ensure the scripts follow the quality standards of elicitation interviews. We outline a research plan to address the shortcomings of our method and to enhance our study with additional future directions. %This plan includes both immediate steps to mitigate current limitations and long-term initiatives to expand the scope and impact of our research.

%\section*{Acknowledgment}
%Acknowledgment

\bibliographystyle{IEEEtran}
\bibliography{gorer}

 \newcommand{\noop}[1]{}
\begin{thebibliography}{10}
\providecommand{\url}[1]{#1}
\csname url@samestyle\endcsname
\providecommand{\newblock}{\relax}
\providecommand{\bibinfo}[2]{#2}
\providecommand{\BIBentrySTDinterwordspacing}{\spaceskip=0pt\relax}
\providecommand{\BIBentryALTinterwordstretchfactor}{4}
\providecommand{\BIBentryALTinterwordspacing}{\spaceskip=\fontdimen2\font plus
\BIBentryALTinterwordstretchfactor\fontdimen3\font minus \fontdimen4\font\relax}
\providecommand{\BIBforeignlanguage}[2]{{%
\expandafter\ifx\csname l@#1\endcsname\relax
\typeout{** WARNING: IEEEtran.bst: No hyphenation pattern has been}%
\typeout{** loaded for the language `#1'. Using the pattern for}%
\typeout{** the default language instead.}%
\else
\language=\csname l@#1\endcsname
\fi
#2}}
\providecommand{\BIBdecl}{\relax}
\BIBdecl

\bibitem{zowghi2005requirements}
D.~Zowghi and C.~Coulin, ``Requirements elicitation: A survey of techniques, approaches, and tools,'' in \emph{Engineering and managing software requirements}.\hskip 1em plus 0.5em minus 0.4em\relax Springer, 2005, pp. 19--46.

\bibitem{ferrari2020sapeer}
A.~Ferrari, P.~Spoletini, M.~Bano, and D.~Zowghi, ``Sapeer and reversesapeer: teaching requirements elicitation interviews with role-playing and role reversal,'' \emph{Requirements Engineering}, vol.~25, no.~4, pp. 417--438, 2020.

\bibitem{bano2019teaching}
M.~Bano, D.~Zowghi, A.~Ferrari, P.~Spoletini, and B.~Donati, ``Teaching requirements elicitation interviews: an empirical study of learning from mistakes,'' \emph{Requirements Engineering}, vol.~24, no.~3, pp. 259--289, 2019.

\bibitem{dalpiaz2021deriving}
F.~Dalpiaz, P.~Gieske, and A.~Sturm, ``On deriving conceptual models from user requirements: An empirical study,'' \emph{Information and Software Technology}, vol. 131, p. 106484, 2021.

\bibitem{ferrari2022requirements}
A.~Ferrari, P.~Spoletini, and S.~Debnath, ``How do requirements evolve during elicitation? an empirical study combining interviews and app store analysis,'' \emph{Requirements Engineering}, vol.~27, no.~4, pp. 489--519, 2022.

\bibitem{sarsa2022automatic}
S.~Sarsa, P.~Denny, A.~Hellas, and J.~Leinonen, ``Automatic generation of programming exercises and code explanations using large language models,'' in \emph{Proceedings of the 2022 ACM Conference on International Computing Education Research-Volume 1}, 2022, pp. 27--43.

\bibitem{bulathwela2023scalable}
S.~Bulathwela, H.~Muse, and E.~Yilmaz, ``Scalable educational question generation with pre-trained language models,'' in \emph{AIED}.\hskip 1em plus 0.5em minus 0.4em\relax Springer, 2023, pp. 327--339.

\bibitem{gorer2023generating}
B.~G{\"o}rer and F.~B. Aydemir, ``Generating requirements elicitation interview scripts with large language models,'' in \emph{2023 IEEE 31st International Requirements Engineering Conference Workshops (REW)}.\hskip 1em plus 0.5em minus 0.4em\relax IEEE, 2023, pp. 44--51.

\bibitem{lo2023trustworthy}
D.~Lo, ``Trustworthy and synergistic artificial intelligence for software engineering: Vision and roadmaps,'' \emph{arXiv:2309.04142}, 2023.

\bibitem{hou2023large}
X.~Hou, Y.~Zhao, Y.~Liu, Z.~Yang, K.~Wang, L.~Li, X.~Luo, D.~Lo, J.~Grundy, and H.~Wang, ``Large language models for software engineering: A systematic literature review,'' \emph{arXiv:2308.10620}, 2023.

\bibitem{siddiq2023exploring}
M.~L. Siddiq, J.~Santos, R.~H. Tanvir, N.~Ulfat, F.~A. Rifat, and V.~C. Lopes, ``Exploring the effectiveness of large language models in generating unit tests,'' \emph{arXiv:2305.00418}, 2023.

\bibitem{fan2023large}
A.~Fan, B.~Gokkaya, M.~Harman, M.~Lyubarskiy, S.~Sengupta, S.~Yoo, and J.~M. Zhang, ``Large language models for software engineering: Survey and open problems,'' \emph{arXiv:2310.03533}, 2023.

\bibitem{kanuka2023exploring}
H.~Kanuka, G.~Koreki, R.~Soga, and K.~Nishikawa, ``Exploring the chatgpt approach for bidirectional traceability problem between design models and code,'' \emph{arXiv:2309.14992}, 2023.

\bibitem{ronanki2023investigating}
K.~Ronanki, C.~Berger, and J.~Horkoff, ``Investigating chatgpt’s potential to assist in requirements elicitation processes,'' in \emph{2023 49th Euromicro Conference on Software Engineering and Advanced Applications (SEAA)}.\hskip 1em plus 0.5em minus 0.4em\relax IEEE, 2023, pp. 354--361.

\bibitem{zhang2023preliminary}
J.~Zhang, Y.~Chen, N.~Niu, and C.~Liu, ``A preliminary evaluation of chatgpt in requirements information retrieval,'' \emph{arXiv:2304.12562}, 2023.

\bibitem{luo2022prcbert}
X.~Luo, Y.~Xue, Z.~Xing, and J.~Sun, ``Prcbert: Prompt learning for requirement classification using bert-based pretrained language models,'' in \emph{Proceedings of the 37th IEEE/ACM International Conference on Automated Software Engineering}, 2022, pp. 1--13.

\bibitem{luitel2023improving}
D.~Luitel, S.~Hassani, and M.~Sabetzadeh, ``Improving requirements completeness: Automated assistance through large language models,'' \emph{arXiv:2308.03784}, 2023.

\bibitem{arora2023advancing}
C.~Arora, J.~Grundy, and M.~Abdelrazek, ``Advancing requirements engineering through generative ai: Assessing the role of llms,'' \emph{arXiv:2310.13976}, 2023.

\bibitem{daun2023chatgpt}
M.~Daun and J.~Brings, ``How chatgpt will change software engineering education,'' in \emph{Proceedings of the 2023 Conference on Innovation and Technology in Computer Science Education V. 1}, 2023, pp. 110--116.

\bibitem{duc2023generative}
A.~N. Duc, T.~L{\o}nnestad, I.~Sundb{\o}, M.~R. Johannessen, V.~Gabriela, S.~U. Ahmed, and R.~El-Gazzar, ``Generative ai in undergraduate information technology education--insights from nine courses,'' \emph{arXiv:2311.10199}, 2023.

\bibitem{macneil2023experiences}
S.~MacNeil, A.~Tran, A.~Hellas, J.~Kim, S.~Sarsa, P.~Denny, S.~Bernstein, and J.~Leinonen, ``Experiences from using code explanations generated by large language models in a web software development e-book,'' in \emph{Proceedings of the 54th ACM Technical Symposium on Computer Science Education V. 1}, 2023, pp. 931--937.

\bibitem{banic2023pair}
B.~Bani{\'c}, M.~Konecki, and M.~Konecki, ``Pair programming education aided by chatgpt,'' in \emph{2023 46th MIPRO ICT and Electronics Convention (MIPRO)}.\hskip 1em plus 0.5em minus 0.4em\relax IEEE, 2023, pp. 911--915.

\bibitem{jalil2023chatgpt}
S.~Jalil, S.~Rafi, T.~D. LaToza, K.~Moran, and W.~Lam, ``Chatgpt and software testing education: Promises \& perils,'' in \emph{2023 IEEE International Conference on Software Testing, Verification and Validation Workshops (ICSTW)}.\hskip 1em plus 0.5em minus 0.4em\relax IEEE, 2023, pp. 4130--4137.

\bibitem{savelka2023large}
J.~Savelka, A.~Agarwal, C.~Bogart, and M.~Sakr, ``Large language models (gpt) struggle to answer multiple-choice questions about code,'' \emph{arXiv:2303.08033}, 2023.

\bibitem{hu2023survey}
L.~Hu, Z.~Liu, Z.~Zhao, L.~Hou, L.~Nie, and J.~Li, ``A survey of knowledge enhanced pre-trained language models,'' \emph{IEEE Transactions on Knowledge and Data Engineering}, 2023.

\bibitem{singhal2023large}
K.~Singhal, S.~Azizi, T.~Tu, S.~S. Mahdavi, J.~Wei, H.~W. Chung, N.~Scales, A.~Tanwani, H.~Cole-Lewis, S.~Pfohl \emph{et~al.}, ``Large language models encode clinical knowledge,'' \emph{Nature}, vol. 620, no. 7972, pp. 172--180, 2023.

\bibitem{wu2023bloomberggpt}
S.~Wu, O.~Irsoy, S.~Lu, V.~Dabravolski, M.~Dredze, S.~Gehrmann, P.~Kambadur, D.~Rosenberg, and G.~Mann, ``Bloomberggpt: A large language model for finance,'' \emph{arXiv:2303.17564}, 2023.

\bibitem{ovadia2023fine}
O.~Ovadia, M.~Brief, M.~Mishaeli, and O.~Elisha, ``Fine-tuning or retrieval? comparing knowledge injection in llms,'' \emph{arXiv:2312.05934}, 2023.

\bibitem{dodge2020fine}
J.~Dodge, G.~Ilharco, R.~Schwartz, A.~Farhadi, H.~Hajishirzi, and N.~Smith, ``Fine-tuning pretrained language models: Weight initializations, data orders, and early stopping,'' \emph{arXiv:2002.06305}, 2020.

\bibitem{chen2021meta}
Y.~Chen, R.~Zhong, S.~Zha, G.~Karypis, and H.~He, ``Meta-learning via language model in-context tuning,'' \emph{arXiv:2110.07814}, 2021.

\bibitem{lewis2020retrieval}
P.~Lewis, E.~Perez, A.~Piktus, F.~Petroni, V.~Karpukhin, N.~Goyal, H.~K{\"u}ttler, M.~Lewis, W.-t. Yih, T.~Rockt{\"a}schel \emph{et~al.}, ``Retrieval-augmented generation for knowledge-intensive nlp tasks,'' \emph{Advances in Neural Information Processing Systems}, vol.~33, pp. 9459--9474, 2020.

\bibitem{mohedas2022use}
I.~Mohedas, S.~R. Daly, R.~P. Loweth, L.~Huynh, G.~L. Cravens, and K.~H. Sienko, ``The use of recommended interviewing practices by novice engineering designers to elicit information during requirements development,'' \emph{Design Science}, vol.~8, p. e16, 2022.

\bibitem{donoghue2000projective}
S.~Donoghue, ``Projective techniques in consumer research,'' \emph{Journal of Consumer Sciences}, vol.~28, 2000.

\bibitem{burnay2014stakeholders}
C.~Burnay, I.~J. Jureta, and S.~Faulkner, ``What stakeholders will or will not say: A theoretical and empirical study of topic importance in requirements engineering elicitation interviews,'' \emph{Information Systems}, vol.~46, pp. 61--81, 2014.

\bibitem{spijkman2023summarization}
T.~Spijkman, X.~de~Bondt, F.~Dalpiaz, and S.~Brinkkemper, ``Summarization of elicitation conversations to locate requirements-relevant information,'' in \emph{REFSQ}.\hskip 1em plus 0.5em minus 0.4em\relax Springer, 2023, pp. 122--139.

\bibitem{openai2023gpt4}
OpenAI, ``Gpt-4 technical report,'' 2023.

\bibitem{park2023longstory}
K.~Park, N.~Yang, and K.~Jung, ``Longstory: Coherent, complete and length controlled long story generation,'' \emph{arXiv:2311.15208}, 2023.

\bibitem{fan2018hierarchical}
A.~Fan, M.~Lewis, and Y.~Dauphin, ``Hierarchical neural story generation,'' \emph{arXiv:1805.04833}, 2018.

\bibitem{rashkin2020plotmachines}
H.~Rashkin, A.~Celikyilmaz, Y.~Choi, and J.~Gao, ``Plotmachines: Outline-conditioned generation with dynamic plot state tracking,'' \emph{arXiv:2004.14967}, 2020.

\bibitem{yang2022re3}
K.~Yang, Y.~Tian, N.~Peng, and D.~Klein, ``Re3: Generating longer stories with recursive reprompting and revision,'' \emph{arXiv:2210.06774}, 2022.

\bibitem{yang2022doc}
K.~Yang, D.~Klein, N.~Peng, and Y.~Tian, ``Doc: Improving long story coherence with detailed outline control,'' \emph{arXiv:2212.10077}, 2022.

\bibitem{celikyilmaz2020evaluation}
A.~Celikyilmaz, E.~Clark, and J.~Gao, ``Evaluation of text generation: A survey,'' \emph{arXiv:2006.14799}, 2020.

\bibitem{zhu2020gruen}
W.~Zhu and S.~Bhat, ``Gruen for evaluating linguistic quality of generated text,'' in \emph{Findings of the Association for Computational Linguistics: EMNLP 2020}, 2020, pp. 94--108.

\bibitem{lending2022rubric}
D.~Lending, J.~D. Ezell, T.~W. Dillon, and J.~May, ``A rubric to evaluate and enhance requirements elicitation interviewing skills,'' \emph{Journal of Information Systems Education}, vol.~33, no.~4, pp. 371--387, 2022.

\bibitem{lee2022personachatgen}
Y.-J. Lee, C.-G. Lim, Y.~Choi, J.-H. Lm, and H.-J. Choi, ``Personachatgen: Generating personalized dialogues using gpt-3,'' in \emph{Proceedings of the 1st Workshop on Customized Chat Grounding Persona and Knowledge}, 2022, pp. 29--48.

\bibitem{10.1145/3241743}
\BIBentryALTinterwordspacing
K.-J. Stol and B.~Fitzgerald, ``The abc of software engineering research,'' \emph{ACM Trans. Softw. Eng. Methodol.}, vol.~27, no.~3, sep 2018. [Online]. Available: \url{https://doi.org/10.1145/3241743}
\BIBentrySTDinterwordspacing

\end{thebibliography}

\end{document}